\documentclass[aps,prl,twocolumn,showpacs,unsortedaddress]{revtex4}

\usepackage[dvips]{graphicx}
\usepackage{amsmath}
\usepackage{amsfonts}
\usepackage{amssymb}
\usepackage[latin1]{inputenc}
\usepackage{ulem}
\usepackage{xcolor}

\begin{document}

\title{Observation of suppression of light scattering induced by dipole-dipole interactions in a cold atomic ensemble}

\author{J. Pellegrino, R. Bourgain, S. Jennewein, Y.R.P. Sortais, and A. Browaeys}
\affiliation{Laboratoire Charles Fabry, Institut d'Optique, CNRS, Univ Paris Sud,
2 Avenue Augustin Fresnel,
91127 Palaiseau cedex, France}
\author{S.D. Jenkins and J. Ruostekoski}
\affiliation{Mathematical Sciences, University of
  Southampton, Southampton SO17 1BJ, United Kingdom}

\date{\today}

\begin{abstract}
We study the emergence of collective scattering in the presence of dipole-dipole interactions when we illuminate a cold cloud of rubidium atoms with
a near-resonant and weak intensity laser. The size of the atomic sample is comparable to the wavelength of light. When we gradually increase the atom
number from 1 to $\sim$ 450, we observe a broadening of the line, a small red shift and, consistently with these, a strong suppression of the
scattered light with respect to the noninteracting atom case. We compare our data to numerical simulations of the optical response, which
include the internal level structure of the atoms.
\end{abstract}
\pacs{42.50.Ct,42.50.Nn,42.25.Fx,32.80.Qk,03.65.Nk}

\maketitle

When resonant emitters, such as atoms, molecules, quantum dots, or meta-material circuits, with a transition at a wavelength $\lambda$, are confined
inside a volume smaller than $\lambda^3$, they are coupled via strong dipole-dipole interactions. In this situation, the response of the ensemble to
near-resonant light is collective and originates from the excitation of collective eigenstates of the system, such as super- and sub- radiant
modes~\cite{Dicke1954,Scully2009,Li2013}. Dipole-dipole interactions  affect the response of the system and the collective scattering of
near-resonant light  differs from the case of an assembly of noninteracting emitters~\cite{Lehmberg1970}. It has even been predicted to be suppressed
for a dense gas of cold two-level atoms~\cite{Bienaime2013}.

Following the recent measurement of the collective Lamb shift~\cite{Friedberg1973} in a Fe layer~\cite{Rohlsberger2010}, in a hot thermal vapor~\cite{Keaveney2012} and in arrays of trapped ions~\cite{Meir2013}, it was pointed out~\cite{Javanainen2013} that  the collective response of interacting emitters is different between ensembles exhibiting inhomogeneous broadening, such as solid state systems 
or thermal vapors, and those free of it, such as cold atomic clouds. In particular, inhomogeneous broadening suppresses the correlations induced by the
interactions between dipoles, leading to the textbook theory of the optical response of continuous media~\cite{Javanainen2013,BornandWolf}. In the absence of broadening, however, this theory fails and should be revisited to include the light-induced correlations~\cite{Morice1995,Ruostekoski1997,Javanainen1999,Kiffner2010,Chomaz2012,Miroshnychenko2013,JenkinsPRA2012, Olmos2013}.
Several recent experiments aiming at studying collective scattering with identical emitters used large and optically thick ensembles of cold atoms~\cite{Bender2010,Chalony2011,Bienaime2010,Balik2013}.
However, the case of a cold atomic ensemble with a size comparable to the optical wavelength has not been studied experimentally, nor has the transition between the well-understood case of scattering by an individual atom~\cite{Bourgain2013a} to collective scattering. In particular, the suppression of light scattering when the number of atoms increases in a regime of collective scattering has never been directly observed.

Here, we study -- both experimentally and theoretically -- the emergence of collective effects in the optical response of a cold atomic sample due to dipole-dipole interactions, as we gradually increase the number of atoms. To do so, we send low-intensity near-resonant laser light onto a cloud containing from 1 to $\sim 450$  cold $^{87}$Rb atoms, with a size comparable to the wavelength of the optical transition at $\lambda=780$~nm. Starting from one atom, we observe a broadening of the line as the number of atoms increases, as well as a small red shift and a strong suppression of the amount of scattered light with respect to the case of noninteracting atoms. We show that this suppression is consistent with the measured broadening and shift. We finally
compare our measurements to a numerical simulation of the response of the system in the low excitation limit, including the internal level
structure of the atoms.

\begin{figure}
\includegraphics[width=8cm]{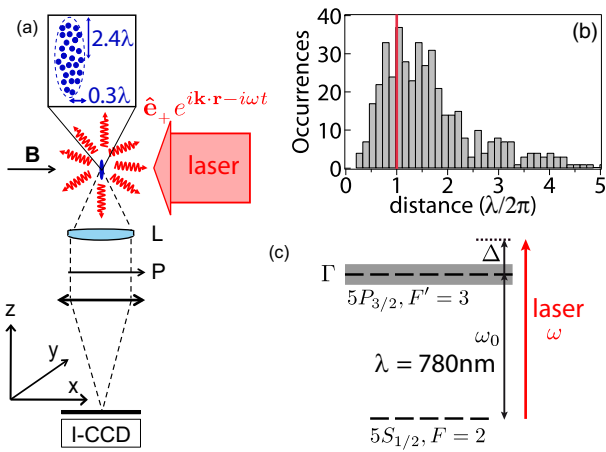}
\caption{(a) Experimental setup. The atoms are initially confined in a microscopic single-beam dipole trap (not shown) (wavelength $957$~nm,  depth $1$~mK, and a waist $1.6~\mu$m, oscillation frequencies $\omega_x=\omega_y=2\pi\times 62$~kHz and $\omega_z= 2\pi\times 8$~kHz). The excitation laser propagates
along the quantization axis $x$, set by a $B\sim1$~G magnetic field. We collect the scattered light along $z$, after a polarizer P oriented at an angle of $55^\circ$ with respect to $x$,  using a lens L with a large numerical aperture (NA$=0.5$) and an image intensifier followed by a CCD camera (I-CCD).
(b) Simulation of the distribution of nearest neighbors for a single stochastic realization of a cloud  of $N=450$ atoms.
(c) Structure of $^{87}$Rb atoms relevant to this work. The excitation light at frequency $\omega$ is near-resonant
with the  transition at
$\lambda=2\pi c/\omega_0=780$~nm.}
\label{Fig:expsetup}
\end{figure}

The suppression of light scattering by resonant dipole-dipole interactions can be understood qualitatively as follows. Consider a laser radiation
with frequency $\omega$ impinging on an ensemble of classical radiating dipoles with resonance frequency $\omega_0={2\pi c/\lambda}$ (see Fig.~\ref{Fig:expsetup}). When the dipoles interact through the dipole-dipole potential
\begin{equation}\label{Eq:dipdip}
V_{j\alpha}^{l\beta}=-V_{\rm dd}\left[ p_{\alpha\beta}(i k r -1) + q_{\alpha\beta}(kr)^2\right]e^{i kr} \ ,
\end{equation}
the system features collective modes with various eigen-frequencies and decay rates. Here, $j$ and $l$ denote two dipoles separated by a distance
$r$, $V_{\rm dd}=3\Gamma/4(k r)^3 $, $k=2\pi/\lambda$, $\Gamma$ is the radiative decay rate in the absence of interactions, and the angular functions
$p_{\alpha\beta}$ and $q_{\alpha\beta}$ depend on the polarizations $\alpha$ and $\beta$  and the relative orientations of the dipoles $j$ and
$l$~\cite{Jackson}~\footnote{Here $p_{\alpha\beta}=\delta_{\alpha\beta}-3({\bf\hat{e}}_\alpha^*\cdot{\bf \hat {r}})({\bf\hat{e}}_\beta\cdot{\bf \hat
{r}})$ and $q_{\alpha\beta}=\delta_{\alpha\beta}-({\bf\hat{e}}_\alpha^*\cdot{\bf \hat {r}})({\bf\hat{e}}_\beta\cdot{\bf \hat {r}})$, with ${\bf \hat
{r}}$ the unit vector along the  interatomic separation and ${\bf\hat{e}}_\sigma$ the unit polarization vector.}. In our experiment, the
geometry of the atomic system and its orientation with respect to the excitation are such that the incident laser best couples to only a few modes
with decay rates $\Gamma_c$ larger than $\Gamma$ (super-radiant modes), leading to a broader excitation spectrum. The excitation rate, and therefore
the amount of scattered light, should thus be reduced by a factor $(\Gamma/\Gamma_{\rm c})^2$. The effect is stronger when the average distance
between dipoles $\langle r \rangle$ is smaller than $\lambda/2\pi$.

To study the collective scattering by an ensemble of atoms coupled via resonant dipole-dipole interactions we use the setup depicted in
Fig.~\ref{Fig:expsetup}(a). We prepare small clouds containing up to 450  atoms at a temperature $\sim100~\mu$K, confined in a microscopic dipole
trap~\cite{Bourgain2013b}, and illuminate them with laser light nearly resonant with the atomic transition at $\lambda=780$~nm. The Doppler width of
the sample ($150$~kHz) is much smaller than the atomic linewidth $\Gamma/2\pi=6$~MHz, making inhomogeneous broadening negligible. The anisotropy of
the trap results in an elongated cloud with calculated root-mean-square thermal sizes $\sigma_{\rho}=0.3\lambda $ and $\sigma_{z}=2.4\lambda$. The
maximal density is $\rho = 2.5\times 10^{14}$ at/${\rm cm}^3$ and the minimal average inter-atomic distance $\langle r \rangle =\rho^{-\frac{1}{3}} =
0.2\lambda$ (Fig.~\ref{Fig:expsetup}b). In this regime, $k\langle r\rangle \sim 1$, leading to $V_{\rm dd}\sim \Gamma$, and the resonant
dipole-dipole interaction will therefore have an effect on the scattering.

Experimentally, we prepare the trapped atoms in the $F=2$ hyperfine manifold with an efficiency better than $95\%$. We then release them in
free space by switching off the trapping light while exciting them with $\sigma_{+}$ polarized light at a frequency $\omega=\omega_0+\Delta$
tuned near the $(5S_{1/2}, F=2)$ to $(5P_{3/2},F'=3)$ transition (see Fig.~\ref{Fig:expsetup}). In this way we avoid extra light-shifts induced by
the trapping beam that would obscure the measurement of small collective shifts and broadening. Also, we choose the intensity saturation $I/I_{\rm
sat}=0.1$ to be in the low excitation limit ($I_{\rm sat}=1.6$ mW/${\rm cm}^2$). We interleave excitation pulses with duration $~125$~ns and
recapture periods in the dipole trap with duration $1~\mu$s. This sequence is repeated 200 times using the same cloud of atoms, in order to improve
the duty cycle of the experiment. Finally, we prepare a new atomic sample and repeat the set of excitation pulses a few hundred times. The scattered
light that we collect in the $z$ direction is therefore the result of an average over many spatial configurations of the atoms. The choice of the
number of pulses (200) is a trade-off between getting a good signal-to-noise ratio and avoiding light-assisted losses~\cite{Fuhrmanek2012} or heating
of the cloud, both of which would lower the density. We checked that both effects do not exceed $5\%$ over the entire set of pulses and that less
than $5\%$ of the atoms are depumped in the $(5S_{1/2}, F=1)$ hyperfine level during the excitation.

\begin{figure}
\includegraphics[width=9cm]{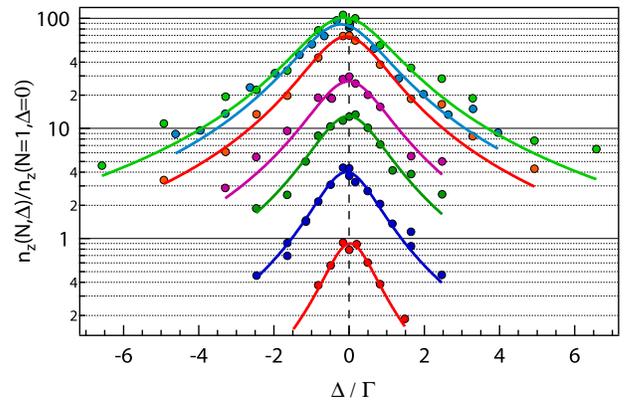}
\caption{Amount of scattered light detected $n_z(N,\Delta)$, versus the detuning $\Delta$ of the excitation light for numbers of atoms $N={1, 5, 20, 50, 200, 325, 450}$ (from bottom to top). The amplitudes of the curves are normalized to the amount of light detected at resonance for a single atom, $n_z(N=1,\Delta=0)$. Solid lines : Lorentzian fits to the data. Typical uncertainties: $10\%$ (vertically) and $20\%$ (horizontally).}
\label{Fig:spectra}
\end{figure}

\begin{figure}
\includegraphics[width=8cm]{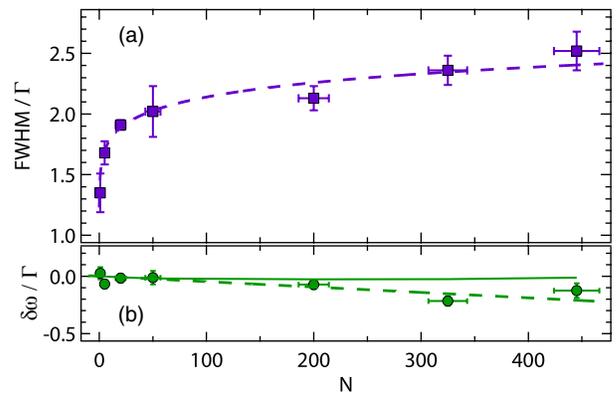}
\caption{(a) FWHM and (b) line shift $\delta\omega$ with respect to the atomic frequency  for $N=1$ atom, in units of $\Gamma$. Filled symbols : data extracted from the Lorentzian fits shown in Fig.~\ref{Fig:spectra}, versus the number of atoms. Dashed lines : phenomenological fits of the FWHM and shift by, respectively, ${\rm \Gamma_{\rm c}}/\Gamma = 1.49(6)\times N^{0.08(1)}$, and $\delta\omega_{\rm c}/\Gamma=47(9)\times10^{-5} N$. The error bars are from the fits of
Fig~\ref{Fig:spectra}. Green solid line: results of the simulation (see text).}
\label{Fig:FWHM_shift}
\end{figure}

\begin{figure}[h!]
\includegraphics[width=8cm]{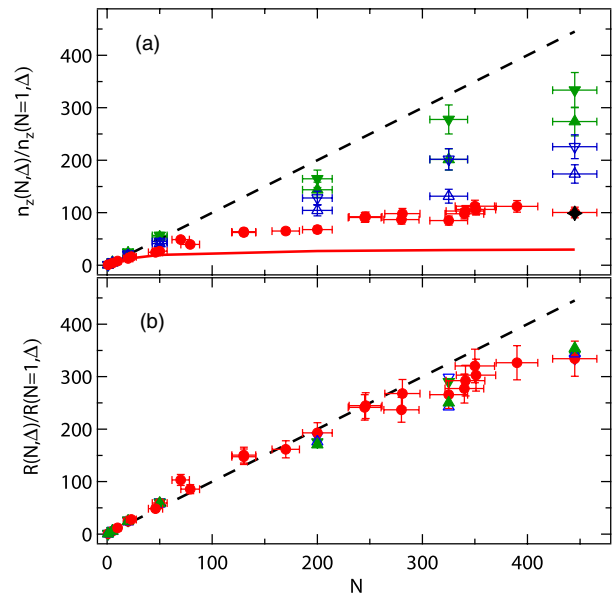}
\caption{(a) Scattered light detected in the $z$ direction, versus the number of atoms, for different detunings of the laser : $\Delta=0$ (red circles), $\pm\Gamma$ (up/down open triangles), and $\Delta=\pm2.5\Gamma$ (up/down filled triangles). The intensity for each atom number is normalized to  the single atom case at the same detuning. Red line: result of the simulation (see text) with widths of the cloud $\sigma_\rho$ and $\sigma_z$. Black diamond: model with widths $2\sigma_\rho$ and $2\sigma_z$. (b) Ratio  $R(N,\Delta)/R(N=1,\Delta)$ versus the number of atoms (see text). Dashed line in (a) and (b): case of noninteracting atoms. }
\label{Fig:suppression_fluo}
\end{figure}

Figure~\ref{Fig:spectra} shows the number of photons $n_z(N,\Delta)$ detected by the I-CCD as a function of the detuning $\Delta$ of the excitation
laser, for various  atom numbers $N$. A Lorentzian fit agrees well with the data for the range of atom numbers explored here. As
expected from the qualitative argument described above, we observe that  the full-width-at-half maximum (FWHM) increases with the number of atoms
(see Fig.~\ref{Fig:FWHM_shift}a), since the interatomic distance then decreases, leading to stronger dipole-dipole interactions. We also measure a
small red shift $\delta\omega$ of the center frequency (Fig.~\ref{Fig:FWHM_shift}b). For $N=1$ atom, the FWHM is $1.35\pm0.15\Gamma$, in agreement
with the short duration of the excitation pulses ($125$~ns), which broadens slightly the resonance. Figure~\ref{Fig:spectra} also shows that the
amount of light scattered in the $z$ direction at resonance does not increase linearly with the number of atoms as one would expect for
noninteracting atoms, but actually increases more slowly. Fig.~\ref{Fig:suppression_fluo}(a) indicates that this is also the case off resonance,
where we plot $n_z(N,\Delta)/n_z(N=1,\Delta)$ for different atom numbers and detunings. For noninteracting atoms this ratio is equal to the number of
atoms $N$ (and is thus independent of the detuning $\Delta$), as we verified by collecting the scattered light after letting the atomic cloud expand
in free space for a sufficiently long time~\cite{Fuhrmanek2010}. By contrast, here we observe that the amount of scattered light is strongly
suppressed on resonance as the number of atoms increases, and that we gradually recover the behavior of noninteracting atoms as we detune the laser
away from resonance.

All the observations reported above can be reproduced by a single functional form:
\begin{equation}\label{Eq:nz}
n_z(N,\Delta)= C\, {N\over \Gamma_{\rm c}(N)^2+4[\Delta-\delta \omega_{\rm c}(N)]^2}\ ,
\end{equation}
where $C$  includes the detection efficiency of the imaging system. 
This is
illustrated in Fig.~\ref{Fig:suppression_fluo}(b): we find that the quantity $R(N,\Delta)/R(N=1,\Delta)$, where
$R(N,\Delta)=n_z(N,\Delta)\times[\Gamma_{\rm c}^2+4(\Delta-\delta \omega_{\rm c})^2]$ and $\Gamma_{\rm c}$ and $\delta\omega_{\rm c}$ are
respectively the phenomenological fits of FWHM and the shift  (see Fig.~\ref{Fig:FWHM_shift}), collapses on a single curve whatever the detuning. For
$N\lesssim300$,  this curve is  linear with $N$ with a slope of $1$, in agreement with Eq.~(\ref{Eq:nz}). It emphasizes that in this regime, the
scattered intensity is suppressed by a factor $(\Gamma/\Gamma_{\rm c})^2$ at resonance, as expected from the qualitative discussion earlier. We note
that this scaling cannot be explained by a model where the suppression would come from an incoherent superposition of the intensities scattered by
each atom with resonant frequencies inhomogeneously distributed over a distribution of width ${\rm FWHM}$: that would lead to a suppression that
would scale as $\Gamma/\Gamma_{\rm c}$ near resonance, instead of the $(\Gamma/\Gamma_{\rm c})^2$ scaling observed here. 
For $N>300$, the departure from the linear law indicates that $C$ depends on the atom number
in this regime, and that the simple Lorentzian form~(\ref{Eq:nz}) becomes inaccurate, 
as also found in the simulation (see below).

We have performed numerical simulations of the collective dynamic response of the atomic sample to near-resonant pulsed light in the low excitation
limit. In this model, each atom, located at position ${\bf r}_j$ ($j=1,\ldots,N$) and with dipole ${\bf d}_j$, is driven by the incident laser field
and by the fields scattered by all the $N-1$ other atoms, i.e. each dipole is coupled to the $N-1$ other dipoles via the resonant interaction of
Eq.~(\ref{Eq:dipdip}). This classical electrodynamics simulation incorporates all the interactions between an ensemble of non saturated discrete
dipoles. This approach has been used to study dielectric media comprising two-level or spatially averaged isotropic electric dipoles
\cite{Javanainen1999,Pierrat2010,JenkinsPRA2012,Chomaz2012,Javanainen2013,Antezza2013} as well as magneto-dielectric circuit resonator systems
\cite{JenkinsPRB2012}. Here, we also incorporate the Zeeman level structure of the atoms \cite{Ruostekoski1997} and the shifts associated to the
presence of the magnetic field. To calculate the dipoles ${\bf d}_j$  in our experimental configuration, we stochastically sample the
positions of the atoms according to a 3-dimensional Gaussian density distribution with root-mean-square sizes given by the thermal sizes of the cloud
along and perpendicular to the trap propagation axis; each atomic position is treated as an independent and identically distributed random variable.
At each realization the $N$ atoms are fixed at positions ${\bf r}_j$ ($j=1,\ldots,N$) and we stochastically sample the magnetic quantum number of the
Zeeman states $m_j$ of each atom $j$. The probability of atom $j$ being in state $|g,m\rangle$ ($m=\pm 2, \pm 1,0$) is the initial population of that
Zeeman state $p_m$ ($0<p_m<1$; $\sum_m p_m=1$). The optical pumping used in the preparation step before the excitation sequence, skews the initial
populations; here we use the values $p_0=p_1=p_2=1/3$ and $p_{-1}=p_{-2}=0$. We write the positive frequency component of the dipole produced by each
atom $j$ that oscillates at the laser frequency as ${\bf d}_j=\mathcal{D}\sum_{\sigma}{\bf\hat{e}}_\sigma C^{(\sigma)}_{m_j}{\cal P}_{j\sigma}$,
where the sum runs over the unit spherical polarization vectors $\sigma=\pm1,0$. The amplitude of the atomic dipole $j$ associated to the optical
transition $|g,m_j\rangle\rightarrow |e,m_j+\sigma\rangle$ is proportional to the reduced dipole matrix element $\mathcal{D}$, the atomic coherence
${\cal P}_{j\sigma}$, and the corresponding Clebsch-Gordan coefficient $C^{(\sigma)}_{m_j}$. The temporal evolution of the coherences is given by the
set of coupled equations
\begin{align}\label{Eq:coupled_dipoles}
  & \dot{\cal P}_{j\alpha}-i\left(\Delta_{j\alpha}+i\Gamma/2\right) {\cal P}_{j\alpha} \nonumber\\
  &= -i\Omega_{j\alpha}(t) -i  \sum_{l\neq j}\sum_{\beta} C^{(\beta)}_{m_l}C^{(\alpha)}_{m_j}{V_{j\alpha}^{l\beta}({\bf r})}\, {\cal P}_{l\beta},
\end{align}
where $\Omega_{j\alpha}(t)$ and $\Delta_{j\alpha} = \omega -\omega_{j\alpha}$ are respectively the time-dependent Rabi frequency and the
detuning of the driving laser with respect to the Zeeman shifted transition of the $\alpha$-polarized atom $j$ with frequency $\omega_{j\alpha}$, and
$\beta=\pm1,0$. Here, we deduce $\Omega_{j\alpha}(t)$ from the experimentally measured temporal profile of the excitation pulse. The last
term in Eq.~(\ref{Eq:coupled_dipoles}) couples the $\alpha$-polarized dipole $j$ to the $\beta$-polarized dipole $l$ separated by ${\bf r}={\bf
r}_{j}-{\bf r}_{l}$ according to Eq.~(\ref{Eq:dipdip}). We have solved Eqs.~(\ref{Eq:coupled_dipoles}) numerically in the presence of a 1G magnetic
field to calculate the light field amplitude that is scattered into the solid angle encompassed by the aspherical lens in the far field. Finally,
accounting for the polarization-sensitive detection scheme, we calculated the measured light intensity.

\begin{figure}
\includegraphics[width=9cm]{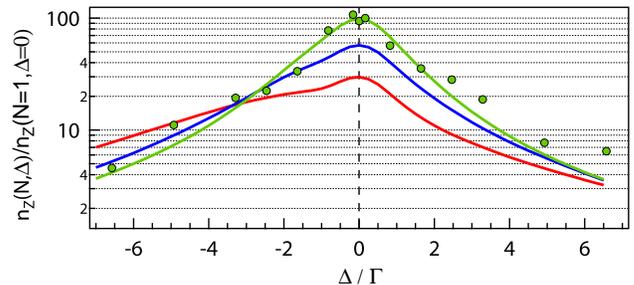}
\caption{Comparison between experiment and theory for the number of detected photons $n_z(N,\Delta)$ (normalized to the single atom case at resonance) for $N=450$. The red, blue and green lines correspond to samples with widths $\sigma_\rho$ and $\sigma_z$ multiplied by a factor  1, 1.44, and 2 respectively.}
\label{Fig:Fig5_comparaison_spectra_volume}
\end{figure}

The simulation predicts that the spectra $n_{z}(N,\Delta)$ should present an increasing broadening and asymmetry
(Fig.~\ref{Fig:Fig5_comparaison_spectra_volume}), a negligible shift (Fig.~\ref{Fig:FWHM_shift}b), as well as a suppression of the scattered light
(Fig.~\ref{Fig:suppression_fluo}a) when the number of atoms increases. These features are in good agreement with our data for $N\lesssim50$. In this
range, the simulated spectra are well fitted by a Lorentzian for $N\lesssim50$, thus justifying our fitting of the data by Eq.~(\ref{Eq:nz}) and the
collapse of the data shown in Fig.~\ref{Fig:suppression_fluo}(b). For $N\gtrsim 50$, the agreement is only qualitative, as the effects are found to
be less pronounced experimentally. We attribute these discrepancies to two possible reasons. Firstly, forces induced by the dipole-dipole
interactions may expel atom pairs with shortest inter-atomic distances, thus breaking down the assumption that atoms have frozen positions during the
sequence of pulses excitations. This is all the more likely as the number of atoms is large, and could explain the evolution of the FWHM in
Fig.~\ref{Fig:FWHM_shift}(a). This effect is hard to check experimentally since the sample is smaller than the diffraction limit of our imaging
system. We found numerically, however, that an increase by a factor 2 in the widths $\sigma_\rho$ and $\sigma_z$ already restores a nearly Lorentzian
profile close to the measured spectra (see Fig.~\ref{Fig:Fig5_comparaison_spectra_volume}), and yields the observed suppression of light scattering
(see Fig.~\ref{Fig:suppression_fluo}a). Secondly, the simulation predicts that the number of detected photons increases by a factor 2 when the
initial distribution of Zeeman state populations varies from $p_0=p_1=p_2=1/3$ to $p_2=1$. For large atom numbers, optical pumping during the set of
excitation pulses may change the distribution of populations, an effect not accounted for in our model.

In conclusion, we have directly measured the suppression of light scattering induced by dipole-dipole interactions in an ensemble of cold atoms
driven by a near-resonant weak laser field and compared it with a time-dependent model of coupled dipoles. The model reproduces the observed
trends. In the future, we plan to investigate to what extent the observed collective scattering  involves beyond-mean-field scattering processes,
i.e. is cooperative in nature. Experimental investigations of the temporal response of the system, and comparisons to the case of a single
atom~\cite{Bourgain2013a}, should also provide insight into the interplay between dipole-dipole interactions and collective scattering.

We acknowledge support from the E.U. through the ERC Starting Grant ARENA, from the Triangle de la Physique, EPSRC, and Leverhulme Trust. We thank P.~Pillet, J.-J.~Greffet, and J.~Javanainen for fruitful discussions.

\end{document}